\documentclass[journal]{IEEEtran}

\usepackage{cite}
\usepackage{graphicx}
\usepackage{gensymb}
\usepackage{mathtools}
\usepackage{color,soul}
\usepackage[colorinlistoftodos,textwidth=10mm]{todonotes} 
\graphicspath{{OBDFleet/}} 
\begin{document}
\title{Design and Implementation of a Wireless OBD II Fleet Management System}

\author{Reza~Malekian,~\IEEEmembership{Member,~IEEE,}
        Ntefeng~Ruth~Moloisane,
        Lakshmi~Nair,~\IEEEmembership{Member,~IEEE,}
        BT(Sunil)~Maharaj,~\IEEEmembership{Member,~IEEE,}
        and~Uche A.K.~Chude-Okonkwo,~\IEEEmembership{Member,~IEEE} 
}

\markboth{Journal of \LaTeX\ Class Files,~Vol.~13, No.~9, September~2014}%
{Shell \MakeLowercase{\textit{et al.}}: Bare Demo of IEEEtran.cls for Journals}
\maketitle

\begin{abstract}
This paper describes the work that has been done in the design and development of a wireless OBD II fleet management system. The system aims to measure speed, distance and fuel consumption of vehicles for tracking and analysis purposes. An OBD II reader is designed to measure speed and mass air flow, from which distance and fuel consumption are also computed. This data is then transmitted via WiFi to a remote server. The system also implements GPS tracking to determine the location of the vehicle. A database management system is implemented at the remote server for the storage and management of transmitted data and a graphical user interface (GUI) is developed for analysing the transmitted data . Various qualification tests are conducted to verify the functionality of the system. The results demonstrate that the system is capable of reading the various parameters, and can successfully process, transmit and display the readings. 
\end{abstract}



\section{Introduction}

\IEEEPARstart{T}{he} on-board diagnostic system (OBD II) is a standard which was developed in the United States of America (USA) in 1996, by the Society of Automotive Engineers (SAE) \cite{SAE2}. This specification was defined for all manufactured vehicles to enable the regulation of vehicle emissions, so as to ensure that the Environmental Protection Agency (EPA) standards are met.

The standard requires that vehicles have a 16-pin OBD II port. Sensor data and diagnostic information from the electronic control unit (ECU) of a vehicle can be measured from this port. 


The development of OBD II also resulted in the development of OBD II scanning tools, known as OBD II readers, which can interface to any vehicle via the 16-pin port. A scanning tool typically requests information from the ECU by sending a message containing a hexadecimal code associated with a specific parameter. These codes are defined by the SAE J1979 standard \cite{SAE3}. The message would then get interpreted according to one of five OBD II signalling protocols. The five OBD II protocols include SAE J1850 (VPW and PWM), ISO 15765, ISO 1941-2 and IS0 142300-4 \cite{Baek}. The ECU finally sends back a hexadecimal code in response. Depending on the specific parameter being measured, the actual measurement can be obtained by simply converting the returned hexadecimal value to decimal or by performing a calculation using a standard formula as defined in \cite{SAE4} for that specific parameter. 

OBD II uses two types of codes to request ECU data. These are Diagnostic Trouble Codes (DTCs) and Parameter Identifiers (PIDs). DTCs are used to diagnose malfunctions in various subsystems of the vehicle and PIDs are used to measure real time parameters. The PID for speed, for instance, is the hexadecimal value $OD$. Vehicle manufactures can define their own PIDs thereby making the on-board system more sophisticated. 

Table \ref{table:1} is a summary of the pin connections of an OBD II reader to the 16- pin connector.

\begin{table}
	\centering
	\caption{OBD II pin connection adapted from \cite{SAE2}}
	\begin{tabular}{|c| c| c|c|} 
		\hline
		Pin Number & Description & Pin Number & Description \\ 
		\hline
		1&	unconnected	&9	&unconnected\\
		2&	J1850 B+	&10 &	J1850 B- \\
		3& unconnected	& 11 &	unconnected\\
		4&	Chassis ground&	12&	unconnected\\
		5&	Signal ground&	13&	unconnected\\
		6&	CAN-H&	14&	CAN-L\\
		7&	ISO K Line	&15&	ISO L\\
		8&		&16&	Battery\\
		\hline
	\end{tabular}
	\label{table:1}
\end{table}

OBD II readers have mostly been utilised for diagnostic purposes; identifying and reporting specific vehicle faults. With the advent of various mobile technologies, wireless communication and the global positioning system (GPS),  the use of OBD II readers for real-time tracking and monitoring has gained popularity, especially in the area of fleet management.

Fleet management is the total management of a company's fleet of vehicles, covering every aspect of the life cycle of a vehicle from procurement to disposal. It is thus important for companies to employ efficient fleet management systems to reduce risks, increase quality of service and improve the operational efficiency of a fleet at minimal cost \cite{Billhardt}. Fleet management also encompasses the analysis of the impact of transportation on the environment. It was reported in \cite{Larue} that approximately 27 $\%$ of the total carbon dioxide ($CO_2$) emissions were as a result of the combustion of fuels from vehicles. 

Vehicle emissions are influenced by driving style and vehicle parameters such as acceleration, speed, distance travelled, and fuel consumption. Speed and distance information in fleet management can be utilised for assessing driver behaviour on the road, prevention of accidents and improved road safety. Fleet management systems can also avail us with a fuel consumption monitoring capability, which can result in reducing running costs and environmental pollution.

Existing works, which are detailed in the next section, have addressed the implementation of OBD II in vehicular systems. Although most of these works integrate OBD II with capabilities for remote position vehicle tracking and system diagnosis, few systems provide the capability for an automated fleet management.

This paper exploits OBD II, GPS, and WiFi technologies to present the design and development of an OBD II-based system for fleet management. The developed OBD II reader can connect to a vehicle's OBD II port and read real-time sensor data from a vehicle's ECU  \cite{Malekian5},\cite{Malekian4}.  The OBD II reader is designed such that it is portable, can be interfaced with any vehicle model and does not interfere with the driving functions while connected. The system provides measurements of speed, distance travelled and fuel consumption. Fuel consumption is computed from the sensor data, since it cannot be directly measured by the OBD II reader. Position or location of the vehicle is also determined by means of a GPS module. Data from the OBD II reader is transmitted to a remote server over a WiFi network.  The use of the WiFi network is considered in this work due to its dominance in the range of technologies for building general purpose wireless networks. A database management system is implemented for the storage and management of transmitted data and a graphical user interface (GUI) is developed for analysing the transmitted data.

The rest of this paper is organised as follows. Section II reviews the related works that were investigated and considered important to this project. Section III provides a brief overview of the modules of the system. Section IV describes the  design and implementation of the system. The observations and results are discussed in Section V, and finally the paper concludes with Section VI.

\section{RELATED WORK}

Various OBD II systems have thus been designed in recent years to solve automotive related problems. Some of these works are discussed below. 

The integration of OBD II and wireless communication technologies was observed in \cite{Baek}, where an OBD II system that measured real time vehicle data was built. The system interfaced with a car's ECU through the OBD II connector. The data received from the ECU was then transmitted to a remote device via Bluetooth, WiFi, or WCDMA in hexadecimal format. The study mainly focused on integrating various wireless communication technologies to connect to various mobile devices. The monitored parameters included vehicle speed and engine revolution per minute (RPM). A flaw in this system is that the received data is not meaningful to a casual user as the hexadecimal data requires decoding.

A system for verification of engine information and diagnosis of engine malfunction using a Bluetooth OBD II scanner was developed in \cite{Kim}. An Android device was used to receive the measured or diagnostic data. The system mainly focused on defining a protocol that enabled transmitting and receiving of OBD II data from multiple sensors simultaneously. This study focused on real-time diagnosis of the engine condition, and data was only made available to the driver of the vehicle.

In \cite{Lin} an OBD system for obtaining engine diagnostic data for air-pollution monitoring was integrated with general packet radio service (GPRS) and GPS technology  \cite{Malekian2}. Three ELM integrated chips (ICs) were used to interpret the different OBD II communication protocols to allow for interfacing with different car types. The system used an RS-232 interface between the OBD reader and a mobile phone and sent data in real-time via GPRS.

The study in \cite{Yang} used an OBD II reader for acquiring real time vehicle parameters from the controller area network (CAN) bus of a hybrid electrical vehicle. The OBD II reader used the ELM 327 IC to interpret the CAN protocol. The data was received wirelessly by an Android device over a Bluetooth network and from the android device, data was sent via GPRS to a remote server.

The impact of driving behaviour on fuel consumption was monitored in \cite{Meseguer} by measuring various parameters such as mass air flow using a Bluetooth OBD II reader. An Android application was used to view the parameters measured for analysis. The measured data was then sent to a web-based remote server. This system exploits the advantage of vehicle on-board systems by using accessible parameters to perform fuel consumption calculations. 

The study in \cite{Zaldivar} implemented an Android-based application that monitored the vehicle via an OBD II interface by measuring the air-bag trigger and G-force experienced by the passenger during a collision, to detect accidents. 

In most of the studies above, the OBD II reader acted as an interface between a mobile device and the ECU of the vehicle. When performing individual vehicle diagnostics or monitoring, these designs would be suitable. However for fleet management systems, solutions that are independent of the type of vehicle and mobile devices in use, are required.

It was also observed from the studies above that real time vehicle parameters can be measured, however there is a limitation in the parameters measurable using standard PIDs defined by OBD II. The distance travelled and fuel consumption, for instance, do not have standard PIDs. These parameters will have to computed from measurable parameters such as mass air flow and speed. There are more traditional methods of measuring speed , as explained in \cite{Geatrix}. These include the use of magnetic sensors, average speed cameras and infra-red devices  \cite{Malekian1} which are normally placed on the road. A disadvantage of using these sensors is that speed can only be measured at specific points and not continuously.

\section{System Overview}
\begin{figure}
	\includegraphics[width=0.5\textwidth]{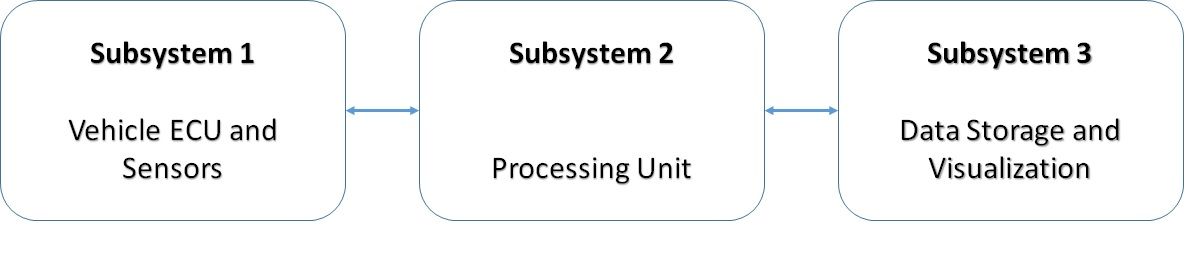}
	\caption{The Three Main Subsystems.}
	\label{fig:12}
\end{figure}

\sethlcolor{green}
A basic overview of the system is given in Figure \ref{fig:12}. 
The ECU of the vehicle is interfaced with various sensors (subsystem 1), from which vehicle parameters can be measured. The OBD II reader (subsystem 2) will be microcontroller based and will thus also be responsible for the control of the overall system. The processed data from subsystem 2 will be transmitted wirelessly to a remote server (subsystem 3) for data storage and display.

\begin{figure}
	\includegraphics[width=0.5\textwidth]{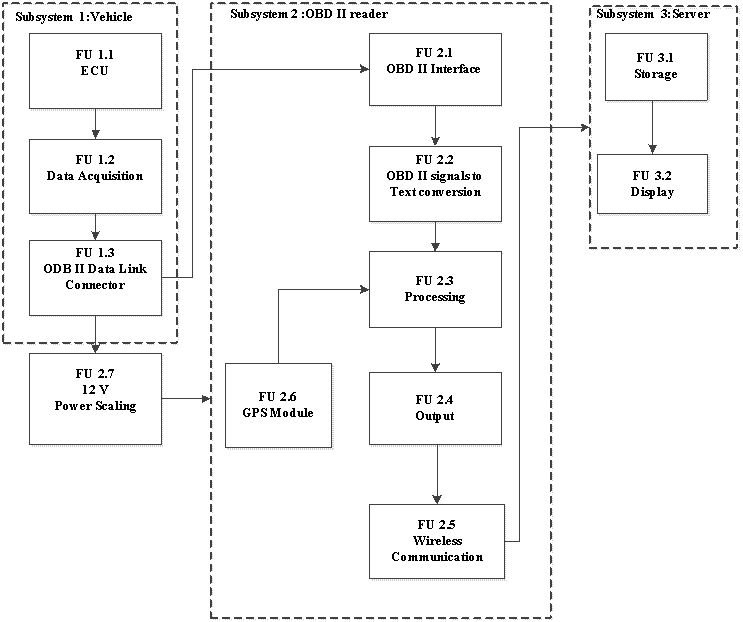}
	\caption{The Functional Units of the Subsystems.}
	\label{fig:13}
\end{figure}

The detailed functions of the subsystems are shown in Figure \ref{fig:13}.

Data acquisition is performed by the vehicle ECU for measurement of the speed, distance travelled and fuel consumption (FU 1.1-1.2). The software to simulate the ECU is designed and implemented on off-the shelf hardware. The data link connector (FU 1.3) is a standard connector in the vehicle to which the OBD II reader is connected. The OBD II protocol interface (FU 2.1) detects and interprets the ECU data according to the implemented OBD II protocol. Conversion of data in text format to voltage levels (FU 2.2) from the processor to the on-board system and vice versa is performed.

PID data is requested and performs the processing of the data received from the ECU via the OBD II protocol interface (FU 2.3). It is also responsible for controlling the GPS (FU 2.6) and the wireless communication modules (FU 2.5). The server is a PC from which a database management system (FU 3.1) and a GUI (FU 3.2) is run. Discrete components and regulators are used (FU 2.7) to scale the 12 V output from the OBD II data link connector down to voltage levels suitable for powering other system components.

\section{System Design}

\subsection{ELM327 Integrated Circuit}
	
The ELM327 is an OBD II interpreter IC. It is a microcontroller designed to automatically interpret all OBD II signalling protocols. It can thus be interfaced with the electronic circuitry required to establish communication with the vehicle ECU via the OBD II port. The protocols implemented in this study were the ISO 15765 (CAN), ISO 9141-2 and ISO 14230-4.
	
An AT Command set predefined for the ELM327 was used to communicate with the IC through RS232. The command set allowed for setting up the IC to change its behaviour so as to suit the requirements of the system. This included setting up the Baud rate for RS232 communication, format of the received data from the ECU and initialisation of the type of OBD II protocol implemented on the vehicle. 
	
A separate command set referred to as OBD commands, was used to communicate with the ECU from the ELM327. Commands from this set consist of only hexadecimal characters as defined in the SAE J1979 standard. Each command from this set is a combination of either a PID or a DTC and a value that indicates a mode of operation. The ELM327 has ten diagnostic modes of operation which are defined in the SAE J1979 standard. The operation of each mode depends on the type of information required from the ECU. Mode one was mainly used in the implementation of the OBD II reader subsystem is for requesting and showing current real time vehicle data and mode two for example is for diagnosing engine malfunctions. The SAEJ1979 standard also defines formulas to be used to decode messages received from the ECU so as to obtain actual parameter measurements and the structure of response message from the ECU.
	
\begin{figure}
	\includegraphics[width=0.45\textwidth]{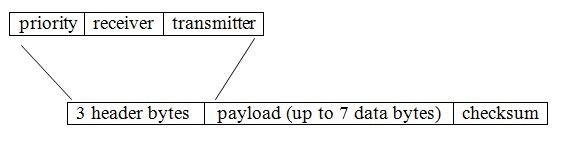}
	\caption{The OBD II Message Format.}
	\label{fig:14}
\end{figure}

OBD commands are typically two bytes long, however they are sent to the vehicle ECU as part of a longer message. The byte structure of this message is shown in Figure \ref{fig:14}.

Messages are assigned priorities which are used to determine the order of sending messages in the event that more than one message is sent simultaneously.The receiver and transmitter bytes are the source and destination addresses. The vehicle ECU is an addressable electrical bus, the source and destination address are thus necessary for use in the address line of the bus.

OBD II commands are encapsulated as part of the payload section which can be as long as 7 bytes. The last field of the message is a checksum which is responsible for detecting errors in messages received from the vehicle ECU. The ISO 9141-2 and ISO 14230-4 standards both employ the same message structure, CAN messages are however slightly different.

\begin{figure}
	\includegraphics[width=0.45\textwidth]{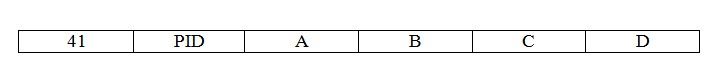}
	\caption{The OBD II Response Message Format.}
	\label{fig:15}
\end{figure}

The structure of response messages from the ECU is shown in Figure \ref{fig:15}. The first byte indicates the mode of operation. The hexadecimal value '1' in the '41' indicates mode 1. If a DTC was sent under mode two then value of the first byte would be '42'. The second byte is the PID which was sent with the request message. Position A to D is the data bytes which contain the requested parameter value from the vehicle. 

\begin{figure}
	\includegraphics[width=0.4\textwidth]{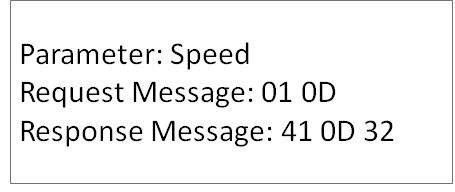}
	\caption{An Example of Request and Response Message Sent for Measuring Speed.}
	\label{fig:16}
\end{figure}

Figure \ref{fig:16} shows an example of a request and a response message sent by the ELM327 and ECU respectively when vehicle speed measurement is requested.  The request message is "01 0D" which indicates that the PID "0D" was sent to the ECU under mode one which is indicated by "01". The response message from the ECU is "41 0D 32" which corresponds to the format shown in Figure \ref{fig:15}. The first two bytes indicate the mode and the PID. The last byte which is "32" in this case, is the data byte A in Figure \ref{fig:15}. The actual speed value is obtained by changing byte A, which is 3216 to decimal. The speed value will thus be 50 Km/h.

Thus the equation are used for calculating Speed is:

\begin{equation}
Speed = ((ByteA)_{16})^{10}
\end{equation}

\begin{figure}
	\includegraphics[width=0.4\textwidth]{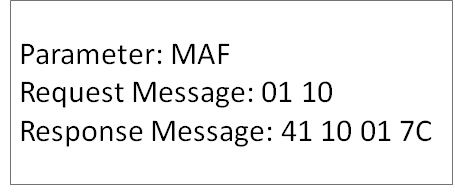}
	\caption{An Example of Request and Response Message Sent for Measuring MAF.}
	\label{fig:38}
\end{figure}

An example of the send and request message for MAF is shown in Figure \ref{fig:38}. The decoding of the data bits is different from the previous example in that the formula used two data bytes, namely A and B.

Computation of the MAF value is done by first converting byte A which is 0116 to a decimal value which is 1. Similarly byte B which is 7C16 works out to be 124.   Thus the equation for calculating MAF is:

\begin{equation}
MAF=\frac{(ByteA_{16})^{10}\times256+(ByteB_{16})^{10}}{100}
\end{equation}

The ELM327 requires a 5 V power source to function correctly. Messages are sent and received using UART. The ELM327 in essence acts as terminal interface. On power up the IC returns a string with the characters "ELM327 v1.4b" followed by a prompt character '$>$'. The prompt character signals that the IC is ready to send OBD II or AT commands.  A baud rate of 38400 bits per second (bps) was used. Establishing communication with the ECU was done by first sending the command ATZ followed by ATSP0 and then finally 0100.The ATZ command resets the IC. It was sent to verify that the IC functions correctly and to check if it was ready to send messages. 

Table \ref{table:2} is a summary of the commands used and sent from the ELM327. All commands sent from the ELM327 were appended with a carriage return character.

\begin{table}
\centering
\caption{The OBDII and AT Commands sent from the ELM 327.}
\begin{tabular}{|c|c|c|} 
	\hline
	 & Command & Description \\ 
	\hline
	1&ATZ&Reset\\
	2&ATSP0&Set protocol to auto\\
	3&ATE0&Echo off\\
	4&ATFE&Forget events\\
	5&ATS0&Print spaces  off\\
	6&0100&Search for set protocol\\
	7&010D&Speed PID\\
	8&0110&MAF PID\\
	\hline
\end{tabular}
\label{table:2}
\end{table}

The OBD II standard does not define standard PIDs for some vehicle parameters such as fuel consumption. A method proposed in [4] was used in this project to calculate the fuel consumption of a vehicle from OBD II. Fuel consumption is a measure of the fuel that a vehicle consumes in litres per kilometre (L/Km). Fuel flow is a measure of the litres of fuel burnt by a vehicle measured in litres per hour (L/h). It can be used to calculate the instantaneous fuel consumption if divided by the current driving speed in kilometres per hour (Km/h). This is shown in equation below.

\begin{equation}
Fuel_{Consumption}=\frac{FuelFlow}{Speed}
\end{equation}

Speed can be obtained from OBD II however even though fuel flow has a defined OBD II PID of 5E, it is not available on most cars. This problem can be bypassed by using the MAF given in grams of air per second (g/s) to calculate fuel consumption. This method takes into account the ratio of the mass of air in grams to one gram of fuel in an engine which is referred to as air to fuel ratio (AFR) and the density of the fuel in grams per cubic decimetre (g/dm\textsuperscript3) or equivalently grams per litre (g/L). The AFR of is 14.7:1 and it's density (D) is 820 g/dm\textsuperscript3. If the speed of the vehicle (V), is given in (Km/h) then the instantaneous fuel consumption can be calculated as in equation below:

\begin{equation}
Fuel_{Consumption}=\frac{MAF}{AFR \times D \times V} \times 3600
\end{equation} 

, where 3600 is a conversion factor from seconds to hours. If the type of fuel used by a vehicle is diesel as opposed to fuel consumption, air to fuel ratio will be 14.5:1 and the density will be 750 g/dm\textsuperscript3. Equation 4 was used in this study to calculate the fuel consumption of a vehicle.

\subsection{Interface Protocols}

\begin{figure}
	\includegraphics[width=0.45\textwidth]{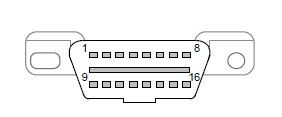}
	\caption{The OBD II Port.}
	\label{fig:17}
\end{figure}

The CAN, ISO 9141-2 and ISO 14230-4 communication protocols were implemented in this study. The output of the interface circuits were connected to the OBD II 16-pin connector shown in Figure \ref{fig:17}.  The pins which were utilised were pin 6 and pin 14 for CAN, pin 7 and 15 for both ISO 9141-2 and ISO 14230-4. Pin 5 and pin 16 were connected to the ground and voltage supply.

\subsubsection{CAN Protocol interface}
	
The CAN standard was developed by a company called Bosch for automotive applications. It was deemed mandatory for all vehicles manufactured from 2008 to implement CAN as the standard OBD II protocol. There are two formats of the CAN protocol with 125 kbps and 500 kbps data transmission rates. The 125 kbps format is referred to as low speed and the 500 kbps as high speed. 
	
\begin{figure}
	\includegraphics[width=0.45\textwidth]{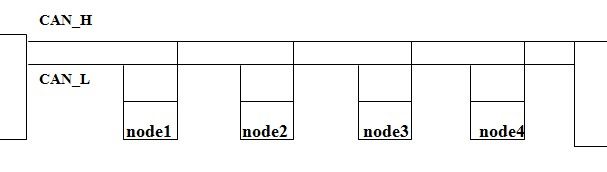}
	\caption{The CAN Data Bus Lines.}
	\label{fig:18}
\end{figure}

The CAN bus standard defines different CAN bus architectures which include a single line and two line bus architecture. Automotive applications, including OBD II mostly employ the double line CAN bus architecture. The double line architecture has transmit and receive lines which connect to different nodes on the bus line, as shown in Figure \ref{fig:18}. Nodes are the different subsystems which can be addressed. The two lines of communication are also referred to as CAN high and CAN low which are characterised by a differential voltage of 5 V and a termination input impedance of 120 $\Omega$.   

\begin{figure}
	\includegraphics[width=0.45\textwidth]{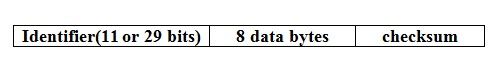}
	\caption{The Structure of a CAN Message.}
	\label{fig:19}
\end{figure}

The structure of CAN messages is different from other OBD II messages. CAN messages also come in two formats depending on the number of identifier bits as shown in Figure \ref{fig:19}. The identifier bits define the message priority and identification of the message stream. A CAN message can either have 11 identifier bits, which is for low speed CAN that operates at a transmission rate of 250 kbps or 29 identifier bits for high speed CAN operating at 500 kbps. The data field which contains the actual data being transmitted on the bus is 8 bytes long and finally checksum bits are defined.

CAN defines states for signals transmitted on the bus line. These signals are just sequences of logic high and logic low voltages also referred to as ones and zeroes respectively. The CAN protocol defines a logical high as a recessive state while a logical low is referred to as dominant state. The definition of these states is based on the value of the differential voltage between CAN high and CAN low data lines. A dominant state is typically when the differential voltage is less that 0 V and a recessive state is when the differential voltage is greater than 1.2 V.

\subsubsection{OBD II ISO 9141-2 interface}

The ISO 9141-2 protocol works on a 10.4 kbps rate. A transmission to the ECU is initialized by sending a 0x33 code at 5 bps. This is referred to as slow initialisation as opposed to fast initialisation used by ISO 14230-4. ISO 9141-2 works on a high 12 V active voltage and a low 0 V passive voltage. It has a single line of communication referred to as K line, where all vehicle ECUs are connected in automotive applications. There may optionally be an additional L line. The K line is accessed through pin 7 of the OBD II connector. The maximum data length is 12 bytes. ISO 14230-4 is similar but differs in data length and initialisation as mentioned previously. The maximum data length is 255 bytes [9].

\begin{figure}
	\includegraphics[width=0.5\textwidth]{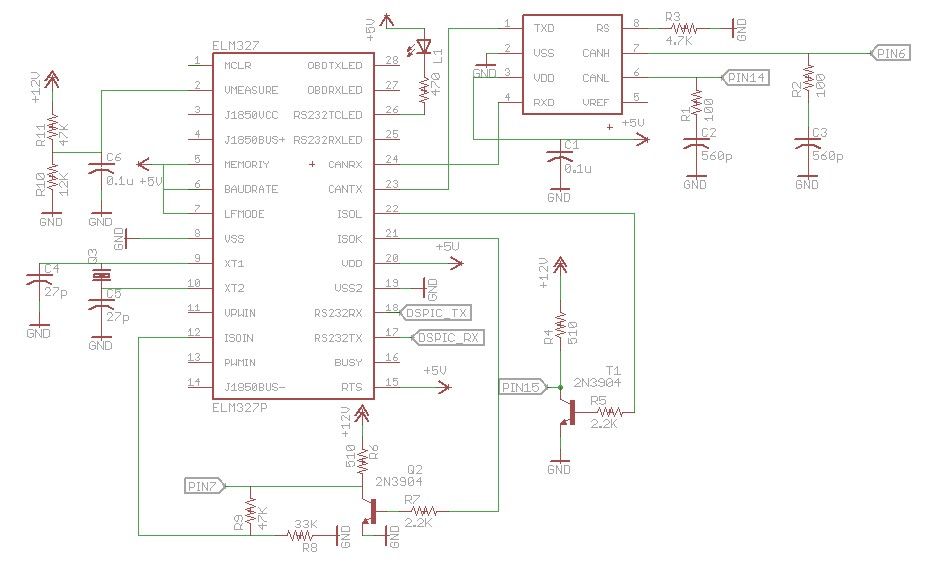}
	\caption{OBD II Communication Protocol Circuitry.}
	\label{fig:20}
\end{figure}

The CAN protocol interface circuit is connected to pin 23 (CAN TX) and pin 24 (CAN RX) of the ELM327 via the MCP2551 which is a CAN transceiver IC. The trans receiver acts as an interface between the vehicle CAN bus and the ELM327 which is responsible for controlling the CAN bus. Controlling the CAN bus entails the transmission and reception of data on the bus. The trans receiver has CAN high and CAN low pins which are connected similarly to the physical vehicle bus via pin 6 pin and pin 14 of the OBD II connector as shown in Figure \ref{fig:20}. Resistors R1 and R2 both of value 100 $\Omega$, are connected to CAN high and CAN low because the ISO 15765-4 requires that a termination impedance between 90 $\Omega$ and 110 $\Omega$ on both CAN high and CAN low lines. Similarly a termination capacitance of 470 pF to 640 pF is required. It is for this reason that the values of C2 and C3 were chosen as 560 pF. Resistor R3 is connected to pin RS of the MCP2551 to control the transition CAN line. A 0.1 $\mu$F capacitor was connected between the positive supply and ground for decoupling and filtering noise in the power line.

The interface circuits for the ISO 9141-2 and ISO 14230-4 protocols are controlled by two NPN transistors which are configured as switches. This is because of the fact that vehicle electronic buses or ECUs that adhere to either of these two standards operate at logical high and logical low voltage levels of 12 V and 0 V respectively. The collector terminals of both transistors are thus connected to the vehicle battery voltage of 12 V via pull up resistors, R4 and R7 of value 510 $\Omega$. The value of the pull up resistor is chosen as specified in the standards of the concerned communication protocols. The operation of the transistor switches when communicating with the ECU is explained further.

When the voltage $V_B$ at the base of the transistor is 0 V, the base current $I_B$ will also be zero. The relationship between the base current and the collector current $I_C$ is given by:

\begin{equation}
I_C=\beta I_{B}
\end{equation}

where $\beta$ is the transistor current gain. If $I_B$ is 0 A then the equation implies that the collector current will also be zero and the transistor will act as an open switch resulting in pin 7 and pin 14 of the OBD II connector being at 12 V. When the transistor is however forward biased with a base voltage of 5 V from the ISO pins of the ELM327, the base current will be:

\begin{equation}
I_B = \frac{V_b - 0.7}{R7}\\
\\=\frac{5-0.7}{2200}\\
\\= 1.95 \\ mA 
\end{equation}

where 0.7 is the transistor breakout voltage.

If the supply voltage is $V_{CC}$, the maximum collector current is given by:

\begin{equation}
I_C = \frac{V_{CC}}{R4}\\
\\=\frac{12}{510}\\
\\= 23.5 \\ mA 
\end{equation}

which would then result in transistor saturating and acting as a closed switch.

The output voltage is given by:

\begin{equation}
V_{out} = V_{CC} - R4I_C\\
\\=12-(510\times0.0235)\\
\\= 0.2 V 
\end{equation}

Calculations performed in Equations 5 to 8 were derived with reference to Q2, however it must be noted that the same reasoning also applies to transistor T1 as the configuration of the two transistor circuits is the same.

Transistor T1 which was connected to the ISOL line is not required for most vehicles as it is only required during the initialisation process of the bus for some vehicles. Data was transmitted and received on the K line of the vehicle bus and was read through pin 12 (ISOIN) of the ELM327 from the output of the transistor. A voltage divider circuit with a resistance R8 of 33 K$\Omega$ and R9 of 47 K$\Omega$ was used to drop down 12 V, the maximum voltage from the transistor output to 5 V to be used by the ELM327. 

The value of the voltage divider output can be justified in the equation below where $V_{out}$ and $V_{in}$ are the voltage divider output and the transistor output respectively.

\begin{equation}
V_{out} = \frac{R8}{R8+R9} \times V_{in}\\
\\=\frac{3300}{3300+4700}\times 12\\
\\= 4.95 V\\
\\\approx 5 V 
\end{equation}

\subsection{Wireless Communication Module}

A WiFi network was established between a Carambola2 WiFi module connected to a PC and one integrated with the OBD II subsystem. The Carambola2 module which was interfaced with the OBD II subsystem was connected to the MCU via RS232. The MCU transmitted the acquired OBD II and GPS data to the module so that it could be sent wirelessly to the remote WiFi module connected to a PC. The Carambola2 operates at a baud rate of 115200 kbps. Figure \ref{fig:30} below illustrates the connectivity between the WiFi module and the MCU.

\begin{figure}
	\includegraphics[width=0.5\textwidth]{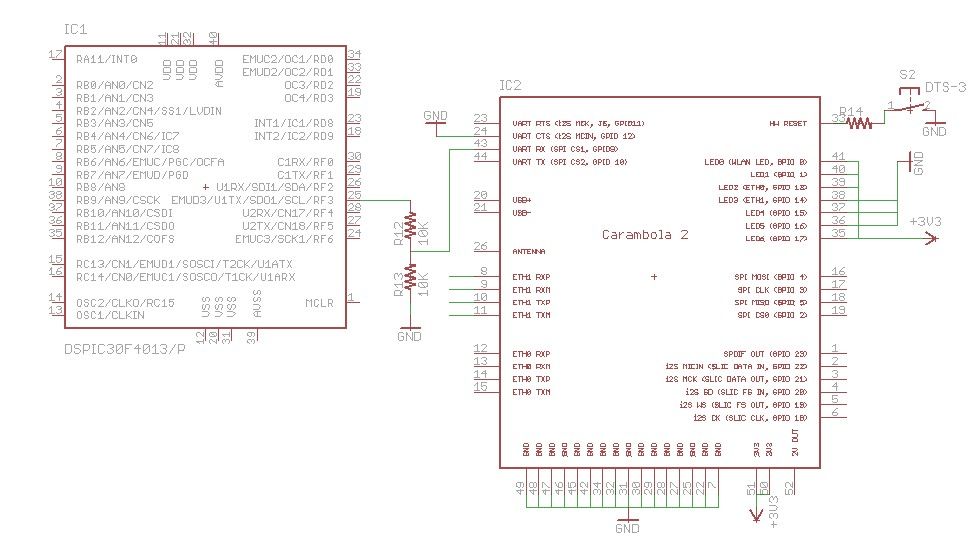}
	\caption{The MCU and WiFi Module Connection Circuit.}
	\label{fig:30}
\end{figure} 

The WiFi module was operated from a 3.3 V supply and the maximum operation voltage of the module's RS232 lines was 2.6 V. Voltage division was hence used to drop the 5 V from the MCU's TX line to 2.5 V suitable for the RX line of the WiFi module. Two equal resistors R12 and R13 of 10 K$\Omega$ were chosen to realise a voltage level of 2.5 V. A level shifting IC such as ADuM1201 could have been used as an alternative to voltage division however voltage division was opted for since it was a cheaper solution and proved to be efficient. The module uses a 2.4 GHz WiFi antenna for WiFi connectivity and scanning of reachable WiFi networks.
	
\begin{figure}
	\includegraphics[width=0.4\textwidth]{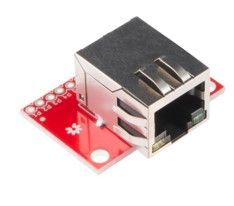}
	\caption{The RJ45 MagJack Breakout Board Which was Interfaced with the Carambola2. }
	\label{fig:31}
\end{figure}

An RJ45 MagJack breakout board shown in Figure \ref{fig:31} was connected to the Ethernet interface pins of the Carambola2 to allow for sharing of source code files from the PC to the module. The board had eight breakout tracks from the pin contacts of the RJ45 to the solder side connections.

The MagJack RJ45 was equipped with a transformer whose primary side was connected to the solder side of the eight pins of the breakout board. The secondary side of the transformer was connected to the pins of the RJ45 port.

\begin{figure}
	\includegraphics[width=0.5\textwidth]{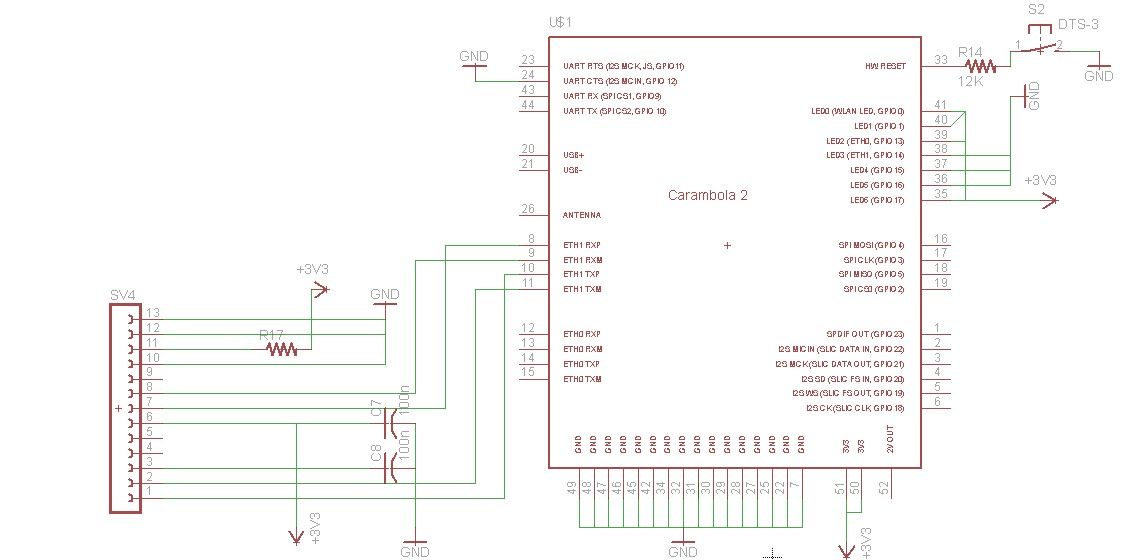}
	\caption{The Connection Between the RJ45 MagJack Breakout and the WiFi Module.}
	\label{fig:32}
\end{figure}

Connections made between the RJ45 breakout board and the LAN interface of the Carambola2 are shown in Figure \ref{fig:32}. The component SV4 was a PCB connector to which the breakout board was connected. Capacitors C7 and C8 of value 100 nF were connected between the supply and ground as suggested in the data sheet for decoupling. A pull down resistor R14 was connected to a switch for physically resetting or rebooting the module.

\subsubsection{(Connecting the Carambola2 Wireless Module)}
The WiFi module was running on Chaos Calmer, a version of the OpenWRT operating system. The Carambola2 is in essence a small computer with powerful features such as those of wireless routers \cite{Malekian3}. Configuring the module for communication over a WiFi network required connecting it first once the module was powered on. This could be done in one of two ways. The first method was through establishing a LAN connection between a PC and the module, then using either Telnet or Secure Shell (SSH) to login. The default IP address of the module's LAN interface was 192.168.1.1 which the PC automatically detected once a successful LAN connection was established.

The second method of accessing the module was through a USB to serial port connection. A terminal program was such as Putty in Windows was used. Connecting to the module via SSH required setting a password which could be done when connected via Telnet or SSH. 
	
The module connected to the PC was configured as an Access Point (AP). An AP acts as a master node or a server which allows for other devices to connect to it via WiFi. Station Point (STA) mode was configured on the module connected to the OBD II subsystem.  STA mode thus enables a device to connect to an AP as a client.

\subsubsection{AP mode configuration}
Once the Carambola module has been connected, the module had to configured for WiFi connection and communication. This was done by setting up three files in the $/etc/config/$ directory of the module. The names of these files were $network$, $wireless$ and $firewall$. 

The $network$ file is where all the networking interfaces to be used for communication are created. The home address of the device was configured as a loopback interface with the IP address 127.0.0.1. The LAN interface was named eth0 and given the IP address 192.168.1.1. This was the IP address used to identify the device when connected to a PC via Ethernet. An interface for WiFi communication was also created and named WiFi The IP address of the WiFi interface was 192.168.6.1. All IP addresses were configured as static addresses with a subnet mask of 255.255.255.0. 

The $wireless$ file is where the wireless interface and the attributes of the WiFi network are configured. These included the name of the WiFi network (SSID), the name of the device, the mode of operation and the communication protocol standard. The device was configured as an access point which could be accessed via the WiFi interface which was initially setup in the $network$ file. The status of the wireless interface was confirmed by issuing the command iwconfig.

The $firewall$ file was setup to enable packets to be forwarded from the LAN interface to the WiFi interface and vice versa.

\subsubsection{STA mode configuration}

Configuring the WiFi module in STA mode is done similarly to the AP mode configurations. The difference was that an interface named wwan with a static IP address of 192.168.6.2 was setup in the $network$ file. The mode of operation and the SSID name were also configured in the $wireless$ file. The same SSID was used, else communication would not have been possible between the modules. The module also allows for enhancing security by setting up a password for the WiFi network.

\subsection{GPS Tracking}

The transmit pin of the GPS module was connected to the receive pin of the microcontroller's second UART module. The module was set to transmit NMEA protocol message strings at a baud rate of 115200 Kbps every second. This was because the WiFi module was also interfaced with the microcontroller’s second UART module and operated at a baud rate of 115200 Kbps. All settings and configurations were done using the U-centre, open source software developed by Ublox. The NMEA protocol defines different messages with a predefined format. One of the NMEA messages is the GLL message type which is string containing information about the longitude and latitude of a specific location.

\begin{figure}
	\includegraphics[width=0.5\textwidth]{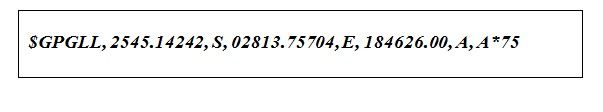}
	\caption{An Example of the GLL Message Transmitted by the GPS Module.}
	\label{fig:34}
\end{figure}

The message string has different fields which are delimited by commas. The fields of the GLL message are shown in Figure \ref{fig:34}.

\subsection{Power Supply}

\begin{figure}
	\includegraphics[width=0.45\textwidth]{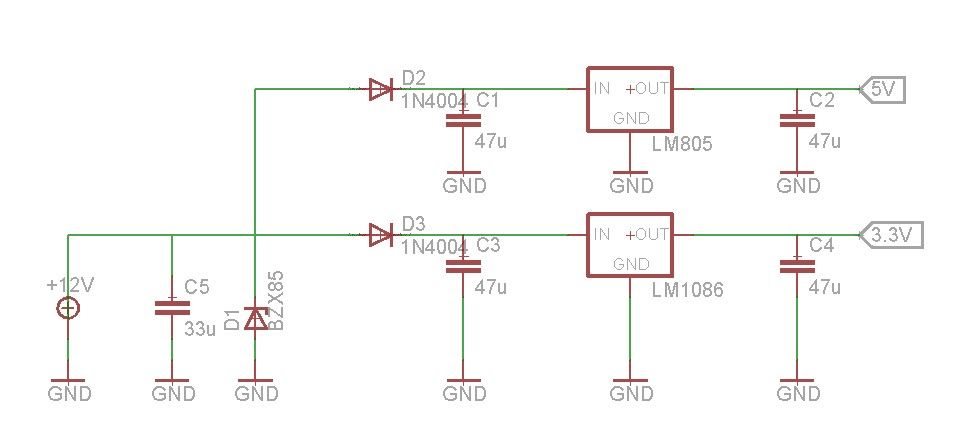}
	\caption{The Power Supply Circuit.}
	\label{fig:35}
\end{figure}

As shown in Figure \ref{fig:35}, the system was powered with a 12 V battery from a vehicle which was regulated down to 5 V the LM805 and 3.3 V using the LM1086 voltage regulators. The 5 V supply was for the ELM327 IC, the MCU and the GPS module. The 3.3 V was for powering the wireless communication module. The two regulators each give a maximum output current of 1.5 A which was sufficient to supply all system components.

Input Capacitors C1, C3 and C5 were used for absorbing power transients and ripples in the circuit. Output capacitors C2 and C4 were also used for the same reason. The values were chosen to be greater than the minimum input or output capacitance as specified in the data sheet A Zener diode D1, with a break down voltage of 12 V was used to protect the system components against overvoltage. This was to limit the voltage from the supply to 12 V as there may be variations when a load is connected. Diodes D2 and D3 with a 0.7 V drop were used to protect supply from any reverse voltage.

\subsection{The Remote Server}
\begin{figure}
	\includegraphics[width=0.45\textwidth]{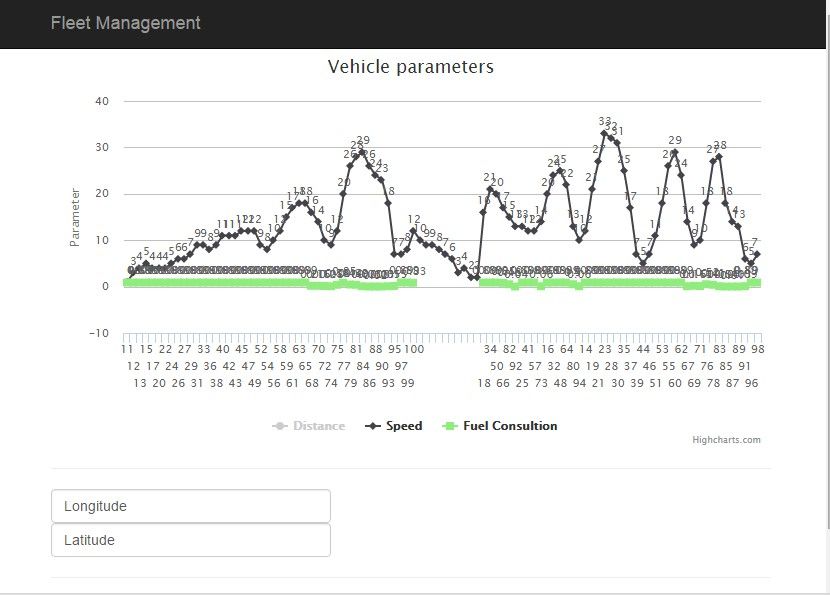}
	\caption{Implementation of the GUI at the Remote Server.}
	\label{fig:36}
\end{figure}

Data from transmitted wireless from the OBD II reader was stored in a database created in SQL. The data was then read from the serial port into a graphical user interface implemented in C\#. The web based model view controller (MVC) platform was used. This allowed for separation of concerns. Figure \ref{fig:36} shows the implementation of the GUI at the remote server. 

\section{Results}

\subsection{Establishing Communication Between the OBD II Reader and the ECU}
Firstly communication between the OBD II reader subsystem and the vehicle ECU needed to be verified. 

\begin{figure}
	\includegraphics[width=0.45\textwidth]{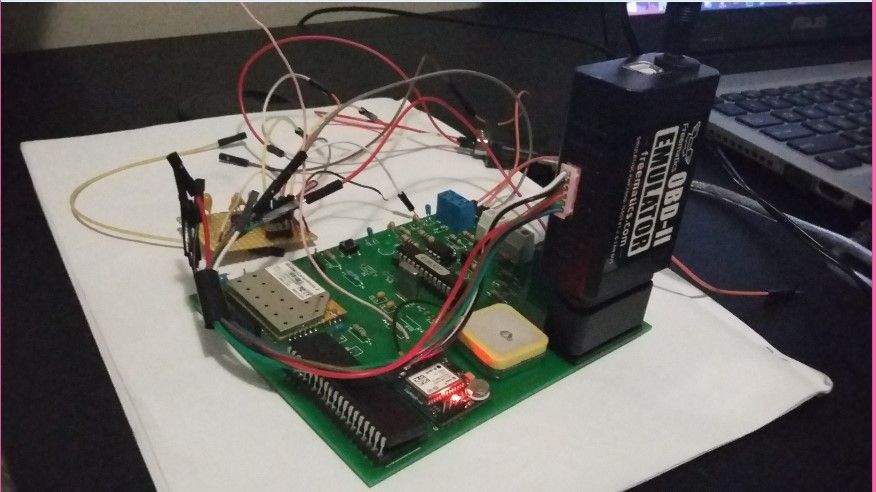}
	\caption{Test Setup of the OBD II Reader Connected to the Emulator.}
	\label{fig:21}
\end{figure}

A Freematics OBD II emulator, as shown in Figure \ref{fig:21}, was used for testing purposes. The emulator implements three OBD II protocols (CAN, ISO 9141-2 and ISO 14230-4. It has an OBD II 16 pin connector similar to an actual OBD II compliant device and was powered from an AC to DC power supply connected to the mains. A USB to serial convertor was used for communication between the PC and the ELM327 IC on the reader. 

\begin{figure}
	\includegraphics[width=0.45\textwidth]{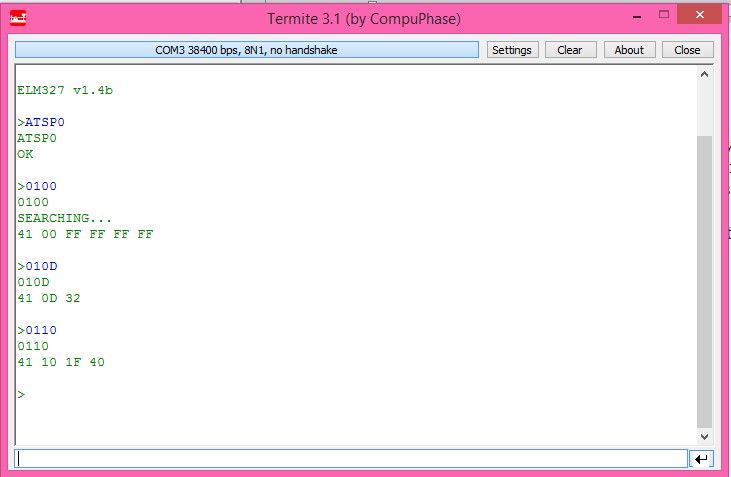}
	\caption{Communication Established Between the Emulator and the Reader.}
	\label{fig:3}
\end{figure}

The speed and MAF parameter values were initially configured on the emulator. AT and OBD II commands were sent from the MCU on the reader to the emulator to initialise communication. The data received from the emulator was then displayed on the terminal program, Termite (which was set at a baud rate of 38400 Kbps), as shown in Figure \ref{fig:3}.

To further verify the designed reader, a Bluetooth OBD II reader was connected to the emulator and an Android application on a mobile phone was used to ensure that the correct data was received.

Commands sent from the MCU were preceded by a prompt character '$>$' and displayed in blue text. The emulator responded successfully to the requests by first echoing the sent command and then responding with the requested data (speed and MAF) shown in green text. 

\subsection{Vehicle Parameter Measurement Over Short Distance,}
The ability and performance of the OBD II reader to measure vehicle parameters: speed, MAF and distance over a short distance was determined.

The designed OBD II reader was connected to the OBD II port of the vehicle. For this study a BMW 125i was used as it implements the CAN communication protocol. A USB to serial convertor cable was connected to a PC running Termite, and the MCU.

\begin{figure}
	\includegraphics[width=0.45\textwidth]{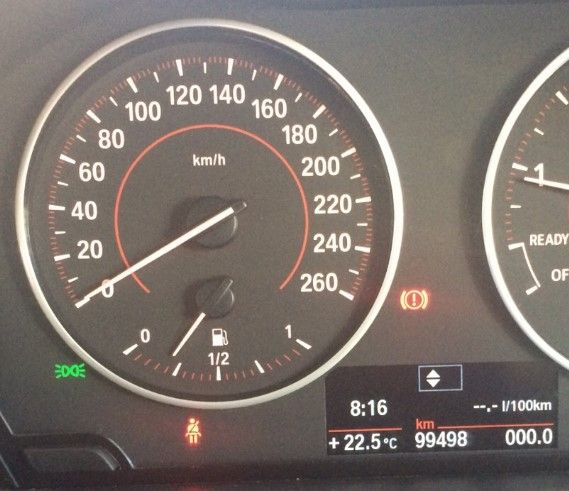}
	\caption{The Vehicle Display.}
	\label{fig:22}
\end{figure}

The vehicle was driven on a straight path for 400 m. The transmitted output was observed on Termite. The on-board computer of the vehicle displayed the instantaneous measurements for speed, distance and fuel consumption. The instantaneous fuel consumption (L/100 km) and distance (km) were shown on the digital display, as in Figure \ref{fig:22}. The speed was observed on a speedometer. A passenger sitting with he driver made the observations in the vehicle.

Samples of all parameters were taken every second for a total duration of 100 s. The instantaneous distance travelled was obtained by multiplying the sampled speed with one second. The resulting value was added to previously obtained results to get the total distance travelled during the course of the trip. 
	
\begin{figure}
	\includegraphics[width=0.45\textwidth]{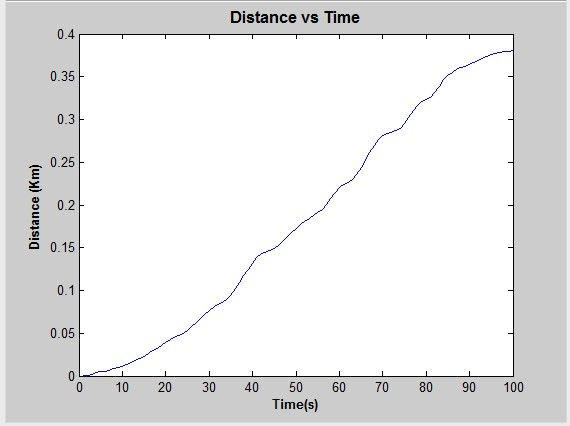}
	\caption{The Relationship Between Distance Covered by the Vehicle and Time.}
	\label{fig:5}
\end{figure}

\begin{figure}
	\includegraphics[width=0.45\textwidth]{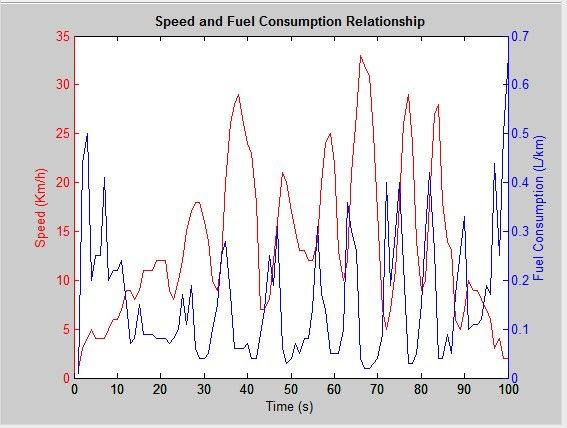}
	\caption{The relationship Between Speed and Fuel Consumption.}
	\label{fig:7}
\end{figure}

The initial and final data points indicate instances when the car started and stopped respectively. The total distance measured by the system was 380 m as shown by the graph in Figure \ref{fig:5}. The maximum speed reached was approximately 34 km/h and the maximum fuel consumed was 0.65 L/km as observed in Figure \ref{fig:7}. It was also noted that more fuel was consumed at lower speeds.


\subsection{Vehicle Parameter Measurement Over Long Distance}
The ability and performance of the OBD II reader to measure vehicle parameters: speed, MAF and distance over longer distances was also determined.

The vehicle was driven over a distance of 9 km from the city to a highway. A total of 419 samples were taken every second for a duration of 450 s. 
	
\begin{figure}
	\includegraphics[width=0.45\textwidth]{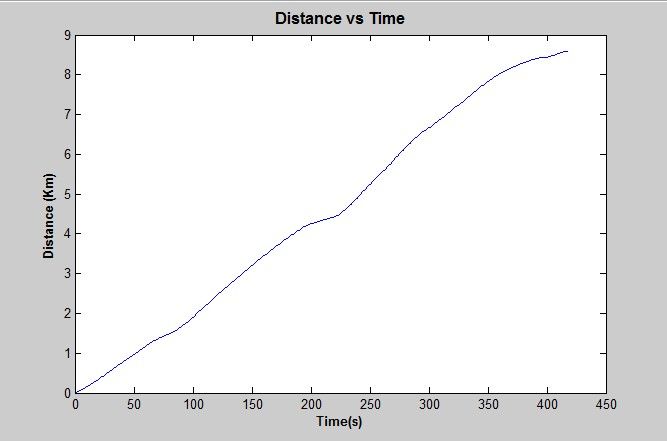}
	\caption{The Relationship Between Distance Covered by the Vehicle and Time.}
	\label{fig:9}
\end{figure}

\begin{figure}
	\includegraphics[width=0.45\textwidth]{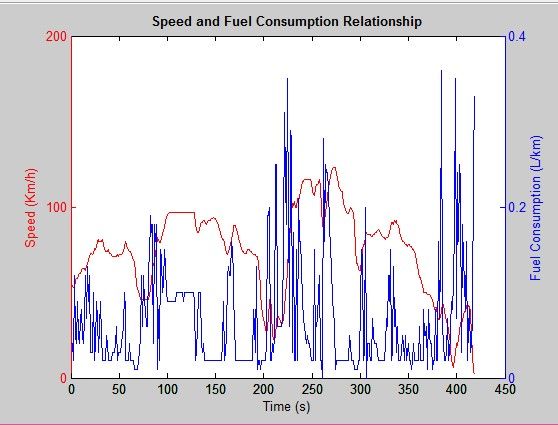}
	\caption{The Relationship Between Speed and Fuel Consumption.}
	\label{fig:11}
\end{figure}

The total distance measured as shown in Figure \ref{fig:9} was 8.5 km. From Figure \ref {fig:11} the maximum speed measured was 120 km/h and the maximum fuel consumption was 0.38 L/km. As noted above the slower the speed the more fuel was consumed.


\subsection{Communication Between the Wireless Modules}
Communication and the range of communication between the two Carambola2 modules is verified.

A remote Carambola module was connected to a PC and the module in the moving vehicle was powered from the vehicle's USB port. Ping commands were constantly issued from the remote module. The success or failure of the ping command was used to determine whether the module in the driving vehicle was out of range or not. 
	
A total communication range of 900 m was determined.

\section{Conclusion}

The testing of the communication between the OBD II reader and the emulator verified that the system could interface with an OBD II complaint vehicle and retrieve sensor measurements. This was further confirmed when the system was tested on a real car and measurements from speed and MAF sensors were successful. 

The design integrates different technologies which can reduce the cost of buying multiple devices with different capabilities. The design also incorporates the measurement of parameters not readily available with most OBD II interfaces, such as GPS tracking, speed and fuel consumption. 

The measurements of the distance travelled by the car showed an error of approximately 5$\%$ which was computed by using the final sample on the distance plot and due to the 1 s sampling period.

It was also observed that when the vehicle's engines were turned off, the system power also shut down. This resulted in the loss of wireless communication and GPS location tracking. 

As was noted above the communication range for the WiFi connection between the remote PC and OBD II reader decreases as distance increases, for future work a GSM system could be considered as part of the communication unit.

Further development can include a battery backup system that supplies power to the rest of the system components when the vehicle's engines are turned off. Thus enabling continued wireless communication and GPS location tracking. The delay resulting from the initialisations of the system could be reduced to improve efficiency of the system. The effect of the delay introduced by initialisations means that the car can only be driven once all initialisations are done, so as to avoid loss of initial parameter measurements. New techniques to reduce this delay need to be developed.



\ifCLASSOPTIONcaptionsoff
  \newpage
\fi



%



\bibliographystyle{IEEEtran}
\bibliography{transportexample}


\begin{IEEEbiography}[{\includegraphics[width=1in,height=1.25in,clip,keepaspectratio]{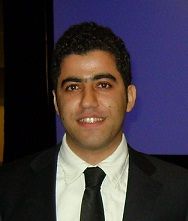}}]{Dr. Reza Malekian}
Reza Malekian (M'10) is currently an Associate Professor with the Department of Electrical, Electronic, and Computer Engineering, University of Pretoria, Pretoria, South Africa. His current research interests include Internet of Things, Sensors and Systems, and mobile communications. Prof. Malekian is also a Chartered Engineer and a Professional Member of the British Computer Society. He is an associate editor for the IEEE Internet of Things Journal.
\end{IEEEbiography}

\begin{IEEEbiography}[{\includegraphics[width=1in,height=1.25in,clip,keepaspectratio]{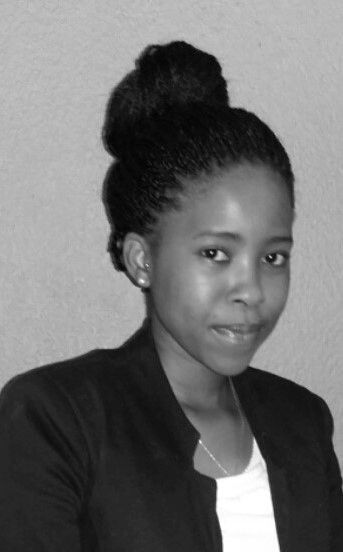}}]{Ntefeng Ruth Moloisane}
received her B.Eng. degree in Computer Engineering from the University of Pretoria, Pretoria, South Africa in 2015. She is currently pursuing her postgraduate studies in the department of Electrical Electronic and Computer Engineering at the University of Pretoria. Her research interests include intelligent transport systems and advanced sensor networks. 
\end{IEEEbiography}

\begin{IEEEbiography}[{\includegraphics[width=1in,height=1.25in,clip,keepaspectratio]{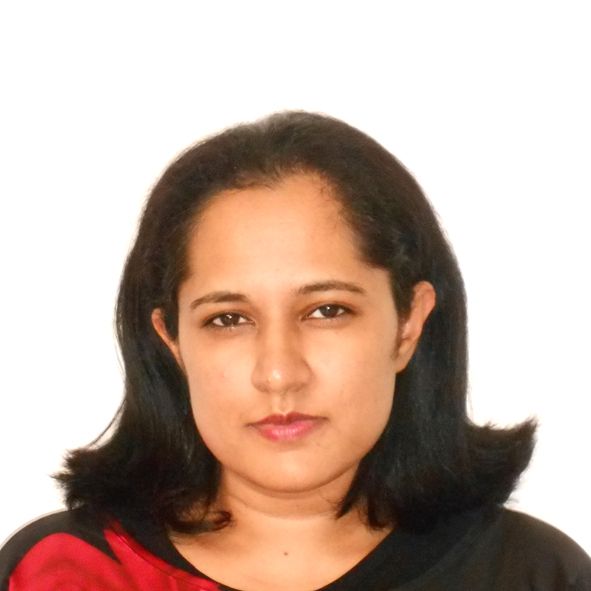}}]{Lakshmi Nair}
received her B.Eng degree in Computer Engineering in 2004 and a M.Eng degree in Electronic Engineering in 2009. She is currently pursuing her Ph.D. degree in Computer Engineering at the University of Pretoria, South Africa. Her areas of interest include wireless sensor networks, intelligent transport systems and smart sensor designs for real-time monitoring applications.
\end{IEEEbiography}

\begin{IEEEbiography}[{\includegraphics[width=1in,height=1.25in,clip,keepaspectratio]{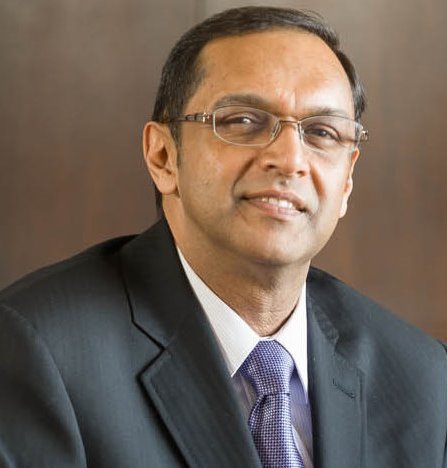}}]{Prof. BT (Sunil) Maharaj }
received his Ph.D. in the area wireless communications from the University of Pretoria, South Africa. Dr Maharaj is a Professor and currently holds the position of Sentech Chair in broadband wireless multimedia communications in the Department of Electrical, Electronic and Computer Engineering at the University of Pretoria. His research interests are in MIMO channel modelling, OFDM-MIMO systems and cognitive radio for rural broadband.
\end{IEEEbiography}

\begin{IEEEbiography}[{\includegraphics[width=1in,height=1.25in,clip,keepaspectratio]{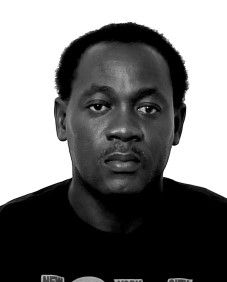}}]{Dr. Uche A.K. Chude-Okonkwo}
	received his Ph.D. in Electrical Engineering from Universiti Teknologi Malaysia in 2010. From 2011 to 2014, he was a Senior Lecturer at the Faculty of Electrical Engineering, Universiti Teknologi Malaysia. He is currently a Senior Research Fellow at the Department of Electrical, Electronics and Computer Engineering, University of Pretoria, South Africa. His current research interests include signal processing and wireless communication.  
\end{IEEEbiography}

\vfill
\vfill
\vfill
\vfill
\vfill
\vfill
\vfill
\vfill
\vfill
\vfill
\vfill
\vfill
\vfill





\end{document}